# Strong Photoluminescence Enhancement of $MoS_2$ through Defect Engineering and Oxygen Bonding

Haiyan Nan[†,&], Zilu Wang[†,&], Wenhui Wang[†], Zheng Liang[‡], Yan Lu[§], Qian Chen[†], Daowei He[∥], Pingheng Tan[§], Feng Miao[⊥], Xinran Wang[∥], Jinlan Wang[†,*] and Zhenhua Ni[†,*]

[†]*Department of Physics, Southeast University, Nanjing 211189, China*

[‡]*Graphene Research and Characterization Center, Taizhou Sunano New Energy Co., Ltd. Taizhou 225300, China*

[§]*State Key Laboratory for Superlattices and Microstructures, Institute of Semiconductors, Chinese Academy of Sciences, Beijing 100083, China*

[∥]*National Laboratory of Solid State Microstructures, Jiangsu Provincial Key Laboratory of Advanced Photonic and Electronic Materials, School of Electronic Science and Engineering, Nanjing University, Nanjing 210093, P. R. China*

[⊥]*National Laboratory of Solid State Microstructures, School of Physics, Nanjing University, Nanjing 210093, P. R. China*

& These authors contribute equally

* Corresponding author: ZH Ni (zhni@seu.edu.cn); JL Wang (jlwang@seu.edu.cn)

# ABSTRACT

We report on a strong photoluminescence (PL) enhancement of monolayer $MoS_2$ through defect engineering and oxygen bonding. Micro- PL and Raman images clearly reveal that the PL enhancement occurs at cracks/defects formed during high temperature vacuum annealing. The PL enhancement at crack/defect sites could be as high as thousands of times after considering the laser spot size. The main reasons of such huge PL enhancement include: (1) the oxygen chemical adsorption induced heavy $p$ doping and the conversion from trion to exciton; (2) the suppression of non-radiative recombination of excitons at defect sites as verified by low temperature PL measurements. First principle calculations reveal a strong binding energy of ~2.395 eV for oxygen molecule adsorbed on an S vacancy of $MoS_2$. The chemical adsorbed oxygen also provides a much more effective charge transfer (0.997 electrons per $O_2$) compared to physical adsorbed oxygen on ideal $MoS_2$ surface. We also demonstrate that the defect engineering and oxygen bonding could be easily realized by oxygen plasma irradiation. X-ray photoelectron spectroscopy further confirms the formation of Mo-O bonding. Our results provide a new route for modulating the optical properties of two dimensional semiconductors. The strong and stable PL from defects sites of $MoS_2$ may have promising applications in optoelectronic devices.

There is a great need for controlling the properties of two dimensional (2D) materials to fulfill the requirements of various applications. For example, modulating the electrical and optical properties of graphene through electrical and magnetic fields,[1] strain,[2] stacking geometry,[3] edge chirality,[4] and defects[5] have been successfully demonstrated.[6] Defects such as vacancies are active centers for molecular adsorption and chemical functionalization, which would provide a great platform for the interplay between 2D materials and various atoms and molecules. Among the mostly investigated 2D layered materials, single and multilayer molybdenum disulphide ($MoS_2$) are semiconductors with direct/indirect bandgap of ~1.2-1.8 eV,[7,8] which makes them promising candidates for optoelectronic applications, such as photodetector,[9,10] photovoltaics,[11-13] and light emitters.[14] However, the photoluminescence (PL) of as-prepared monolayer $MoS_2$ has lagged behind expectation for a high quality direct bandgap semiconductor.[8,15] Previous work has shown that the weak PL is mostly due to formation of negative charged excitons (also named as negative trions) in the naturally *n*-doped $MoS_2$.[16] The switch from trion to exciton in $MoS_2$ *via* electrical gating and molecular adsorptions (*e.g.* $O_2$, $H_2O$, TCNQ) can dramatically enhance the PL of $MoS_2$.[16-19] Structural defects, such as vacancies, dislocations, grain boundaries, and edges have been observed both in pristine/as-grown $MoS_2$[20-22] and electron beam/plasma irradiated samples.[23,24] The proper utilization of these defects to improve the optical properties of $MoS_2$ is highly desirable.

In this work, we present the defect engineering of monolayer $MoS_2$ and the strong oxygen bonding on the defect sites. The PL intensities at the defect sites could be enhanced by at least thousands of times, as observed by high resolution micro-PL images. The oxygen adsorbed on defect site has very strong bonding energy, would introduce heavy $p$ doping in MoS$_2$ and hence a conversion from trion to exciton. The excitons at defects sites are dominated by radiative recombination at room temperature, resulting in a high PL quantum efficiency. We also show that the defect engineering could be easily realized by oxygen plasma irradiation.

**RESULTS AND DISCUSSION**

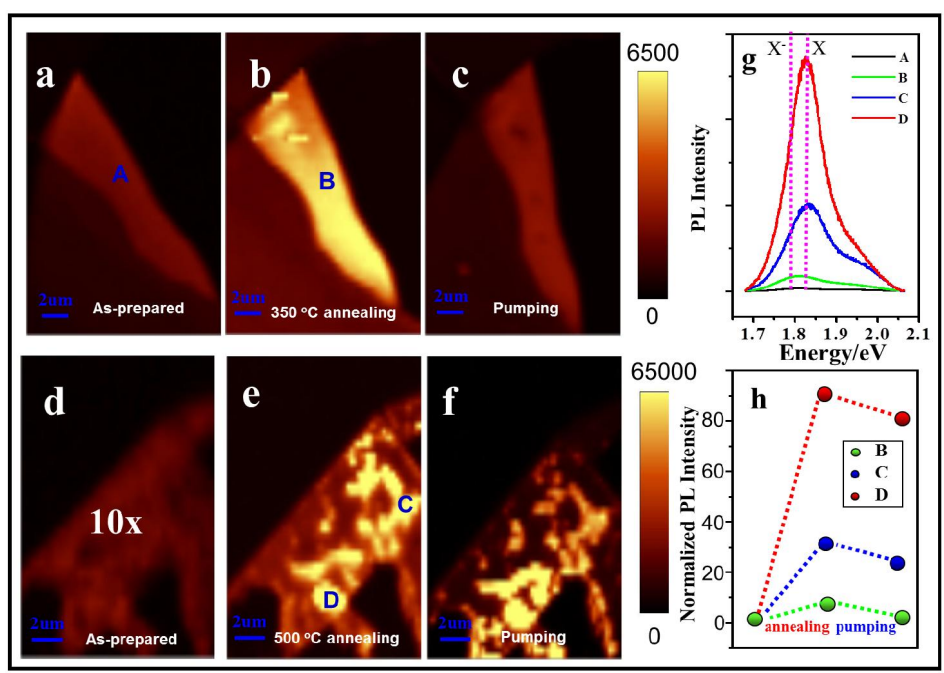

**Figure 1.** PL intensity images of monolayer MoS$_2$: (a) as prepared, (b) annealed in vacuum for one hour at 350 °C, (c) after pumped down to 0.1 Pa. The images have same color bar; PL intensity images of another monolayer MoS$_2$: (d) as prepared, (e) annealed in vacuum for one hour at 500 °C, (f) after pumped down to 0.1 Pa. The images have same color bar; (g) PL spectra taken from locations A-D in the images; (h) The change of normalized PL intensities (as compared to the original values) of locations B-D throughout the annealing and pumping process.

Figure 1a shows PL intensity image of an as-prepared monolayer $MoS_2$, which is very uniform cross the whole sample. The PL spectrum from location A is shown by the black curve in Figure 1g, with a relatively weak peak located at ~1.79 eV, corresponding to the direct bandgap emission. The PL intensity is enhanced by 6 times (Figure 1b) after the sample is annealed for one hour at 350 °C in vacuum (0.1 Pa) and exposed to air ambient. At the same time, the PL peak blueshifts to ~1.81 eV, as shown by the green curve in Figure 1g. These phenomena have been reported and explained by the physical adsorption of $O_2$ and $H_2O$ molecules on $MoS_2$ (*p* type doping, with $O_2$ as the dominant contributor).[18] The as-prepared $MoS_2$ is normally *n* doped, due to the presence of defects or substrate unintentional doping.[20,25] In such case, the photoexcited electron-hole pairs would bind with excess electrons to form negative trions (labeled as $X^-$, with lower energy), instead of neutral excitons (labeled as X, with higher energy). The depletion of excess electrons by $O_2/H_2O$ adsorption can therefore switch the dominant PL process from trion recombination to exciton recombination. The physical adsorbed molecules are not stable due to the weak binding energy and can be easily removed by vacuum pumping. Figure 1c shows that the PL intensity almost reduces to its original value after pumped down to 0.1 Pa. On the other hand, the results are quite different for sample annealed at higher temperature. Figure 1d-1f show PL intensity images of another monolayer $MoS_2$ after similar annealing and pumping processes, but with a higher annealing temperature of 500 °C. As shown in Figure 1e, the PL image becomes highly inhomogeneous after annealing. In some locations, *e.g.* C and D, the PL intensities are dramatically increased to 30 and 89 times, respectively, as compared to its original values. Furthermore, the PL intensities from those locations only drop slightly (to 22 and 80 times, respectively) after the pumping process (Figure 1f and 1h), and they are very stable after laser irradiation of ~30 min with power of ~0.5 mW. This suggests that in these locations, there is a strong interaction between $MoS_2$ and adsorbed molecules, possibly with a chemical adsorption.

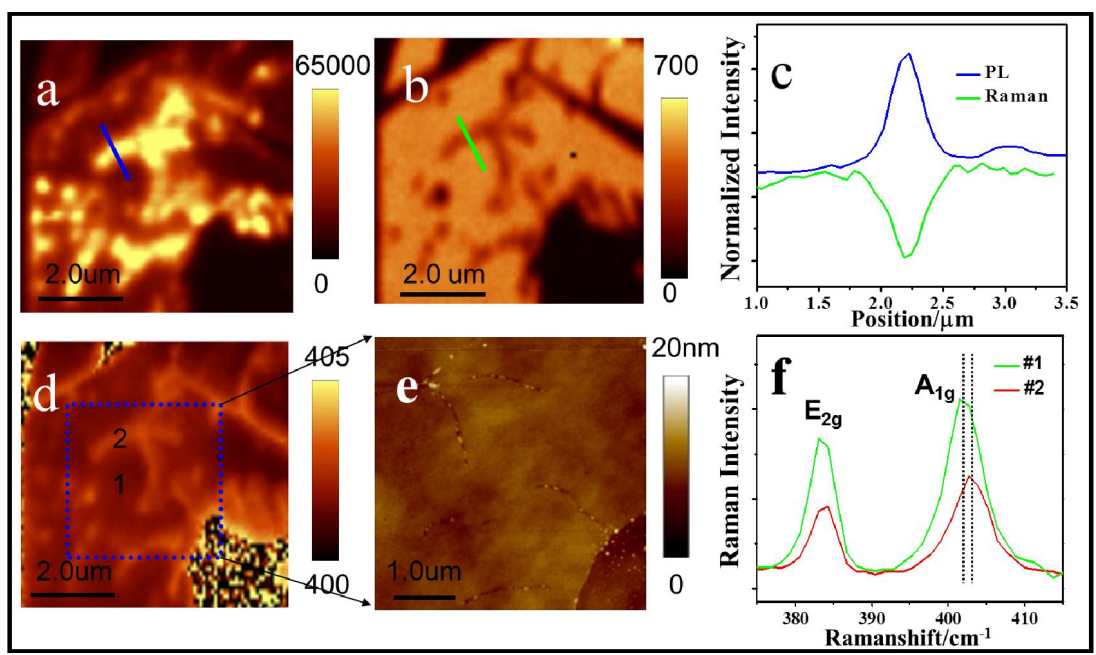

**Figure 2.** PL (a) and Raman $A_{1g}$ peak (b) intensity images of monolayer $MoS_2$ after annealed for one hour in 500°C. (c) Normalized intensity profiles of a crack in (a) and (b); (d) Raman image of the $A_{1g}$ peak frequency; (e) AFM image corresponding to the blue square in (d); (f) Raman spectra taken from locations #1 and #2 in (d).

In order to have a more detailed investigation, high resolution micro- PL and Raman images, as well as atomic force microscope (AFM) image are taken from a selected region of the second sample. Figure 2a and 2b present PL and Raman ($A_{1g}$ peak) intensity images, where the fine structures can be distinguished with a high spatial resolution of ~300 nm. As can be seen, there is a very good correlation between the PL and Raman intensity images, with higher PL intensities in lower Raman intensities regions. AFM investigation in Figure 2e further reveals that these regions are exactly corresponding to the cracks in $MoS_2$, which are formed during the high temperature annealing. The origin of the formation of cracks will be discussed later. We also observe a clear blueshift of the $MoS_2$ $A_{1g}$ peak in the cracked regions as shown in Figure 2d. Raman spectroscopy is a powerful tool to study the properties of $MoS_2$.[26,27] A $p$/$n$ type doping would cause a blueshift/redshift of $A_{1g}$ peak of $MoS_2$ due to the change of electron-phonon interactions, while the $E_{2g}$ peak frequency is almost unaffected.[28,29] The $A_{1g}$ peak of cracked region is slightly blueshifted by ~1 cm$^{-1}$ as compared to other regions (Figure 2f), which suggests that there is a $p$ type doping. It could be understood as following: Mo and S atoms at the cracks of $MoS_2$ contain numerous of dangling bonds, which can be considered as a lot of defects and are very active centers for molecular adsorption. Therefore, $O_2$ can adsorb on the cracks/defects with much stronger binding energy as well as introduce $p$ type doping, as compared to the ideal $MoS_2$ surface. This is also supported by first principle calculations in the later section. The $p$ type doping would introduce a conversion from trion to exciton emissions, as verified by the blueshift of PL peak in Figure 1g (from ~1.79 eV ($X^-$) of as-prepared sample to ~1.83 eV (X) of cracked regions). The intensity profiles of PL and Raman images across a crack are shown in Figure 1c. Both of these two curves can be well fitted by Gaussian functions. As the width of crack is only around 30 nm, the width of Raman intensity profile of ~300 nm is only limited by the spatial resolution of the system. The PL intensity profile has similar width of ~304 nm which suggests that the PL signals are mainly from the cracked regions. In viewing of this, the PL enhancement at cracks/defects regions should be re-estimated by considering the laser spot size. In a normal microscope setup, the PL signals are collected from the whole regions of spot size, *i.e.* ~300 nm in our case. The PL signals from cracks/defects with nanometer sizes are ~30 times even stronger compared to the original PL intensity collected from the whole ~300 nm spot size regions! By considering the cracks as two one-dimensional edges with effective widths of several nanometers, this would give a PL enhancement of at least thousands of times at the crack/defect sites. Such a huge PL enhancement cannot be simply explained by the switch of PL process from trion recombination to exciton recombination, which can only enhance the PL intensities by several to tens of times.[16-19]

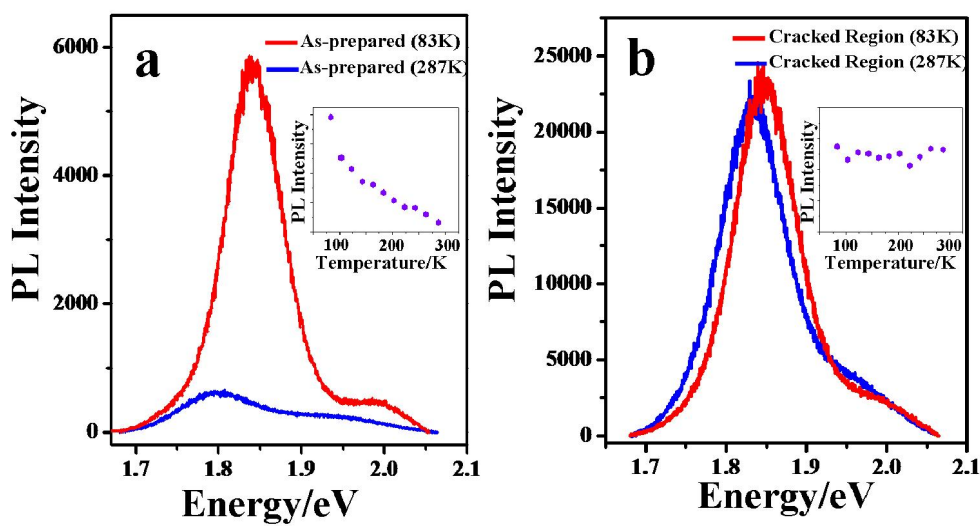

**Figure 3.** PL spectra of (a) as-prepared monolayer $MoS_2$ and (b) cracked regions in annealed MoS$_2$, taken at 83K and 287K. The changes of PL intensities with the change of temperatures (287-83K) are shown in the insets.

We suggest that this is related to the high PL quantum efficiency of excitons at defect sites. Low temperature PL measurements are carried out and shown in Figure 3. The PL spectra from as-prepared monolayer MoS$_2$ as well as cracked regions (similar to Locations C and D in Figure 1e) are collected at temperatures from 83 to 287K in a low temperature Linkam stage (with 1 °C temperature accuracy and stability). The temperature dependence of PL intensity can be expressed by: [30,31]

$$I_{PL}(T) = I_0 * k_{rad}(T)/(k_{rad}(T) + k_{nonrad}(T)) \qquad (1)$$

Where $I_0$ is the maximum PL intensity at very low temperature, and $k_{rad}(T)$ and $k_{nonrad}(T)$ are the temperature dependent radiative and non-radiative recombination rates. The non-radiative recombination rates $k_{nonrad}$ contrains the rates of defect trapping $k_{defect}$ and electron relaxation within the conduction and valence band $k_{relax}$.[8] In most of the materials, the PL intensities as well as quantum efficiency dramatically decrease as the temperature increases, due to the thermally activated non-radiative recombination. The PL intensity of as-prepared MoS$_2$ is reduced to ~10% with the increase of temperature from 83 to 287K (Figure 3a and inset), because of non-radiative recombination. Surprisingly, the PL intensity of cracked regions is almost unchanged with the change of temperature (Figure 3b and inset). According to Equation 1, the non-radiative recombination rate $k_{nonrad}$ of the excitons at cracked regions is almost negligible, and the PL process is dominated by radiative recombination.[32] This could be the main reason of huge PL enhancement at defect/crack sites. The excitons in lower dimensional systems (in our case, the excitons may localize at the defect sites) generally have much larger binding energy,[33,34] which may suppress the thermally activated non-radiative recombination including defect trapping and result in a very high PL quantum efficiency. Previous work also found that the thermal activation energy (temperature) for non-radiative recombination increases with the decrease of quantum well thickness, due to the stronger quantum confinement effect.[35] Further work on the time-resolved PL experiments to investigate the recombination process of these localized excitons would be very interesting. Recently, a strong PL enhancement at $WS_2$ edges has also been reported, and was attributed to the exciton accumulation or localized excitons near lattice defects (different structure and chemical composition of the platelet edges).[36] The PL enhancement did not happen at $WS_2$ edges created *via* mechanical scratch, which suggest that it might be due to the W-O bonding formed during the growth of $WS_2$ at high temperature, since the raw material is $WO_3$.

We have also checked the origin of crack formation and oxygen bonding by changing the vacuum pressure in the annealing chamber. As can be seen in Figure S1(a), annealing in $10^{-4}$ Pa in 600 °C does not introduce oxygen bonding, indicated by the recovery of PL intensity to its original value after vacuum pumping. There are no cracks formed either. However, if we change the vacuum pressure to 100 Pa (which means there are certain amount of oxygen in the chamber at 600 °C), a stronger PL enhancement is observed and the PL intensity only decreases by ~50% after vacuum pumping. This suggests that oxygen in chamber play an important role in the formation of cracks and oxygen bonding. We also anneal the sample in ~300 °C in air, and observe the formation of triangle pits in MoS$_2$, which is similar to the results reported in ref 37. There is also a strong PL enhancement as can be seen in Figure S1(b). Therefore, it is suggested that the cracks in our sample are formed by reaction of O$_2$ (due to the not very high vacuum condition) with MoS$_2$ at high temperature, and followed by the chemical bonding of oxygen molecule at the crack edges at high temperature.

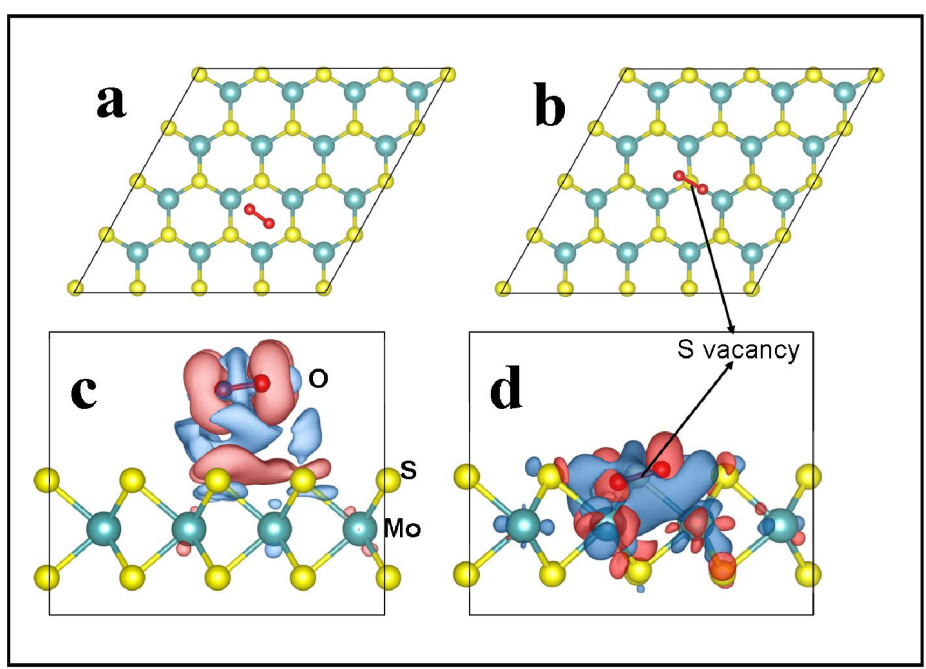

**Figure 4**. Relaxed configuration and charge density difference of O$_2$ molecule physisorbed on perfect monolayer MoS$_2$ (a,c) and chemisorbed on defective monolayer MoS$_2$ containing a mono-sulfur vacancy (b,d). The positive and negative charges are shown in red and blue colors, respectively. Isosurface values are $3\times10^{-4}$ and $1\times10^{-2}$ e/Å$^3$ for (c) and (d), respectively.

The binding energy as well as charge transfer between O$_2$ molecule and MoS$_2$ is calculated by first-principles method. Previous experimental and theoretical results have shown that mono-sulfur vacancies are the most stable defective structures [21] and therefore our simulation is focused on a mono-sulfur vacancy on MoS$_2$ lattice. The structures of O$_2$ molecule adsorbed on an ideal 4×4 supercell of monolayer MoS$_2$ and the same supercell in the presence of an S vacancy are shown in Figure 4a and 4b, respectively. The binding energy between O$_2$ molecule and ideal MoS$_2$ is only 0.102 eV and therefore can be considered as physical adsorption. In contrast, the binding energy of O$_2$ molecule on an S vacancy of MoS$_2$ is 2.395 eV. This suggests that O$_2$ molecule on S vacancy of MoS$_2$ can be considered as chemi-sorbed, and therefore is very stable under vacuum pumping. We have calculated the energy barrier between physical and chemical adsorption of oxygen molecule on MoS$_2$ vacancy, and found a reaction barrier of ~1.05 eV, which can be easily overcome under high temperature annealing conditions. Details can be found in supporting information Figure S2. The charge transfer between an O$_2$ molecule and ideal MoS$_2$ (physical adsorption) is only 0.021e (from MoS$_2$ to oxygen) while that between an O$_2$ molecule and S vacancy (chemical adsorption) reaches 0.997e (from MoS$_2$ to oxygen). The increase of charge transfer between oxygen and MoS$_2$ with S vacancy is also one of the reasons of huge PL enhancement at cracked regions. It should be noted that the charge transfer is strongly localized at the defect site (Figure 4d), which means that the defect can be treated as a hole localization center.[38] The binding of free electrons or trions with such localized holes would form localized excitons, which are very stable and could even avoid non-radiative recombination. This would be the main reason of huge PL enhancement at defect/crack sites of MoS$_2$.

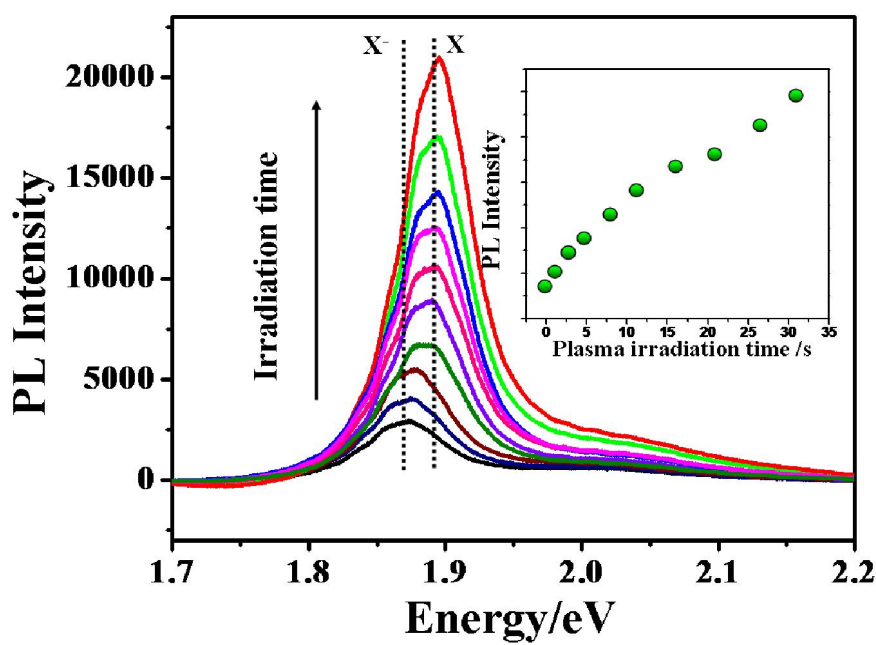

**Figure 5.** PL spectra of monolayer MoS$_2$ after oxygen plasma irradiation with different durations. The change of PL intensities with plasma irradiation times is shown in the inset.

Since the PL enhancement occurs at the defect sites, controlling the defect concentration (especially S vacancies) and introducing oxygen adsorption can be a promising method for manipulating the optical properties of MoS$_2$. Here we adopt mild oxygen plasma (13.56 MHz, 5W, 5 Pa) irradiation to controllably introduce defects in MoS$_2$. It has been reported that plasma irradiation can easily introduce S vacancies in MoS$_2$.[23,24] At the same time, the oxygen ions are more reactive and can easily interact with MoS$_2$ at the defect sites. As shown in Figure 5 and inset, the PL intensity increases gradually with the increase of plasma irradiation time. By careful control of experimental conditions, the PL enhancement could be as high as 100 times. As the power of oxygen plasma is very weak, it can hardly introduce damages to $MoS_2$, which is verified by the almost unchanged Raman line shapes (see Figure S3). The PL intensities of plasma irradiated $MoS_2$ only drops by ~10-20% under vacuum pumping and even after one hour annealing at 400 °C, indicating a strong oxygen chemical adsorption.

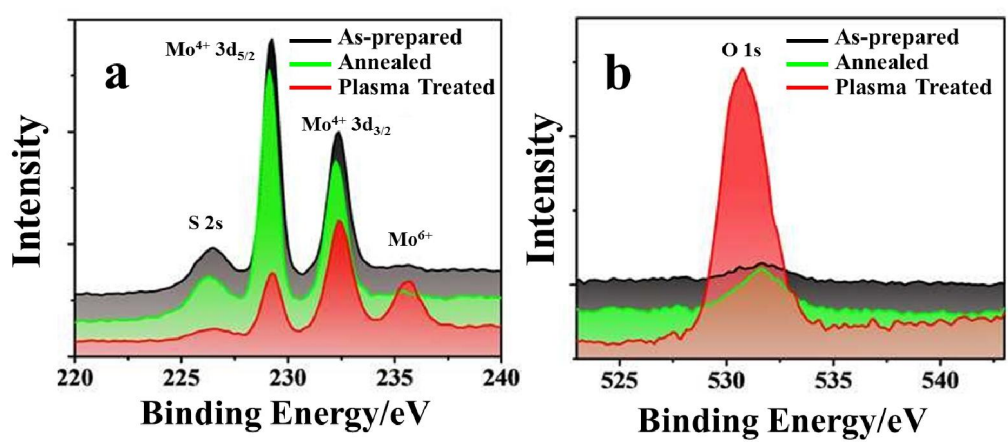

**Figure 6.** XPS spectra of (a) Mo 3d and S 2s core levels (b) O 1s binding energy region measured from as-prepared, annealed at 350 °C for one hour, and oxygen-plasma treated $MoS_2$.

X-ray photoelectron spectroscopy (XPS) is used to study the change of $MoS_2$ structure after annealing and oxygen plasma treatment. Here, a bulk sample is used due to the experimental limitation. The XPS spectra of as-prepared, 350 °C annealed, and oxygen plasma treated (13.56 MHz, 5W, 5 Pa, 30s) $MoS_2$ are shown in Figure 6. All of the samples contain the characteristic peaks of $MoS_2$, with a doublet $Mo^{4+}$ $3d_{3/2}$ and $Mo^{4+}$ $3d_{5/2}$ at ~232.6 and ~229.4 eV, and an S 2s peak at binding energies of

~226.4 eV (Figure 6a).[39] The XPS spectra of as-prepared and 350 °C annealed sample are similar, but with a slightly stronger O 1S peak located at ~532 eV for annealed sample (Figure 6b). This should be due to the physical adsorbed oxygen molecules. In the oxygen plasma treated sample, a well-pronounced peak at ~236.2 eV shows up. This peak is attributed to the $Mo^{6+}$ state, which is also a doublet but with one of the peak overlaps with $Mo^{4+}$ $3d_{3/2}$ at ~233 eV.[40] At the same time, $Mo^{4+}$ and S 2s peaks are reduced. In addition, the intensity of O 1S peak increases dramatically after oxygen plasma treatment (Figure 6b). The appearance of higher oxidized $Mo^{6+}$ states and the strongly increased oxygen concentration in plasma treated $MoS_2$ are clear signatures of the formation of Mo-O bonds.[40] There might be $MoO_3$ formed after oxygen plasma treatment. However, $MoO_3$ does not present obvious PL signals, and therefore does not contribute to the strong PL enhancement.

Finally, it should be noted that vacancy defects in $MoS_2$ would introduce localized donor states inside the bandgap of MoS2 and low-carrier-density transport is dominated by hopping *via* these localized gap states.[20] However, it seems that certain amount of defects in $MoS_2$ generated by thermal annealing and plasma treatment would not degrade its optical properties, while provide the possibility of defect engineering through chemical bonding with oxygen or other types of molecule.

**CONCLUSION**

To summarize, we have achieved a strong PL enhancement of $MoS_2$ through defect engineering and oxygen bonding. The micro- PL and Raman images clearly reveal that the PL enhancement occurs at the defect sites and can be as high as thousands fold after considering the laser spot size. Such a huge PL enhancement is attributed to oxygen bonding induced heavy *p* doping as well as high quantum efficiency of excitons localized at defect sites. Finally, we adopt oxygen plasma to controllably introduce defects and oxygen bonding in $MoS_2$, results in a controllable manipulation of PL. Our results provide a new route to modulate the optical properties of $MoS_2$. The strong PL of $MoS_2$ achieved by defect engineering would be very promising for optoelectronic applications, *e.g.* light emitting devices. [41]

**EXPERIMENTAL METHODS**

**Sample preparation, thermal annealing and vacuum pumping.** Monolayer $MoS_2$ flakes are exfoliated from bulk crystals (SPI Supplies) and transferred to Si wafer with a 300 nm $SiO_2$ capping layer. The thickness of monolayer $MoS_2$ is confirmed by Raman spectra, optical contrast and AFM.[15] The samples are annealed at the temperature of 350 or 500°C for one hour in a quartz tube with pressure of 0.1 Pa. After annealing, the samples are taken out and studied by PL/Raman measurements. The pumping treatment is performed in a homemade vacuum chamber with a pressure 0.1 Pa, which can be put under the Raman microscope.

**Micro-PL/Raman, AFM, and XPS Measurements.** Micro- Rman/PL measurements are carried out using a Witec alpha 300R confocal Raman system. A 100× objective lens with numerical aperture (NA) of 0.9 is used in the Raman and PL experiments, and the spot size of a 532 nm laser is estimated to be ~300 nm using a scanning edge method.[42] The excitation source is 532 nm laser (2.33 eV) with a laser power below 0.5 mW to avoid laser-induced sample damage. For PL and Raman images, the sample is placed on an *x-y* piezostage and scanned under the illumination of laser with a step size of 100 nm. The PL and Raman spectra from every spot of the sample are recorded. Data analysis is done by using WITec Project 2.10 software. AFM is carried out using a Nanoscape 8.15 system with tapping mode. XPS (ULVAC-PHI PHI5000) equipped with a monochromatic Al K α (1486.7 eV) X-ray source is used to scrutinize $MoS_2$ surface.

**Density functional theory calculations.** Spin-polarized density functional theory (DFT) calculations are performed by employing the van der Waals density functional (vdW-DF) [43] within the framework of plane wave pseudopotential method, which is implemented in the Vienna ab initio simulation package (VASP).[44] The optB86b-vdW exchange functional[45] is used for the vdW correction since its estimated lattice constant for $MoS_2$ is closest to the experimental value (3.16 Å) in our calculations. A large spacing of 15 Å between adjacent single layers is used to prevent interlayer interactions. A plane-wave basis set with kinetic energy cutoff of 400 eV is used. The Brillouin zone is sampled by 5×5×1 k-point meshes within Monkhorst-Pack scheme.[46] All atomic positions are optimized until the maximum Hellmann-Feynman forces acting on each atom is less than 0.01 eV/Å.

**ACKNOWLEDGMENTS**

This work is supported by NSFC (11104026, 61376104, 21173040, 21373045,

61325020, 61261160499), Program for New Century Excellent Talents in University (NCET-11-0094), NBRP (2013CBA01600, 2011CB302004 and 2010CB923401), the open research fund of State Key Laboratory of Advanced Optical Communication Systems and Networks, SJTU, China and Natural Science Foundation of Jiangsu Province (BK2011585, BE2011159, BK20130016). We would like to thank Weiwei Zhao, Yumeng You, Da Zhan, Chuanhong Jin and Xing Wu for their helps on this work and computational resources at SEU and National Supercomputing Center in Tianjin.

# Supplementary material for "Strong Photoluminescence Enhancement of MoS$_2$ through Defect Engineering and Oxygen Bonding"

*Haiyan Nan, Zilu Wang, Wenhui Wang, Zheng Liang, Yan Lu, Qian Chen, Daowei He, Pingheng Tan, Feng Miao, Xinran Wang, Jinlan Wang and Zhenhua Ni*

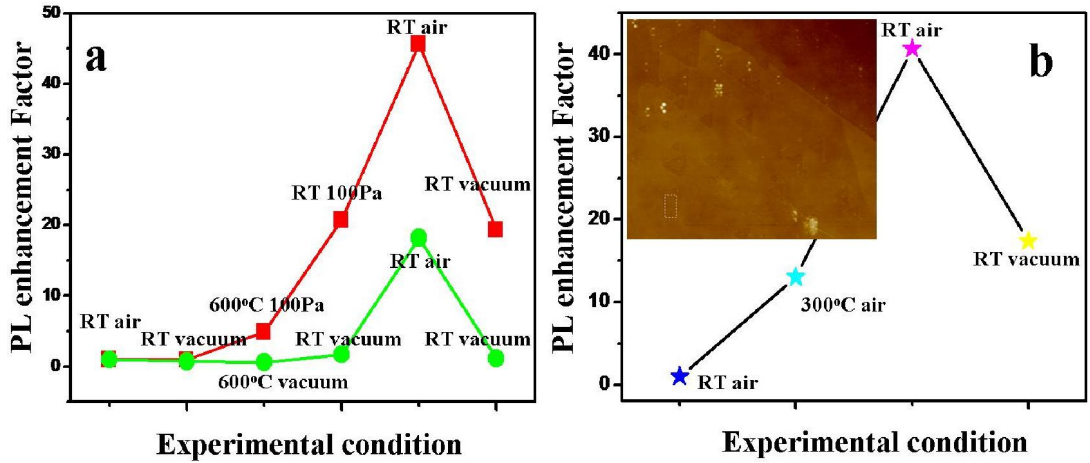

**Figure S1**. (a) The change of PL intensity of MoS$_2$ in different experimental conditions. The difference of green and red curves is the vacuum pressure at high temperature annealing: $10^{-4}$ Pa for green curve and 100 Pa for red curve. (b) The change of PL intensity of MoS$_2$ after annealing in air for half hour at 300 $^o$C. Inset is the AFM image of the sample after annealing. The triangle pits can be clearly observed.

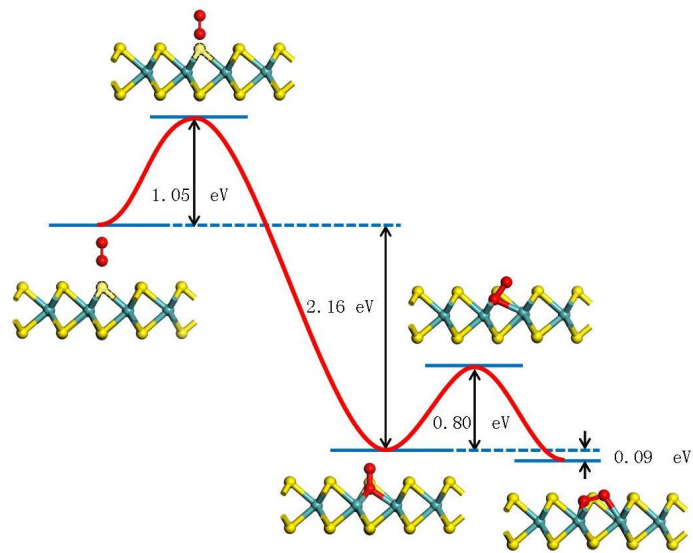

**Figure S2**. The reaction barrier between physical and chemical adsorption of oxygen molecule on MoS$_2$ vacancy.

We calculated the reaction barrier between physical and chemical adsorption of oxygen molecule on MoS₂ vacancy using the climbing-image nudged elastic band (cNEB) method.[1] The reaction path consists of two steps, as demonstrated in Figure S2. In the first step, the physisorbed O₂ approaches MoS₂ sheet and the oxygen atom at the lower end forms chemical bonding with the unsaturated Mo atoms. This process need to overcome a barrier of 1.05 eV with an energy drop of 2.16 eV. In the second step, it takes 0.8 eV to accommodate the O₂ from the standing configuration to the lying-down configuration, which has an energy advantage of 0.09 eV. The barrier of the entire process (1.05 eV) implies that the transition between O₂ physisorption and chemisorption may occur at a moderate temperature of about 440 K according to transition state theory ($t \sim 10^{12} \times \exp(-E_b/kT) \approx 1$, where $E_b = 1.05 eV$ ),[2] which is feasible under our experimental conditions.

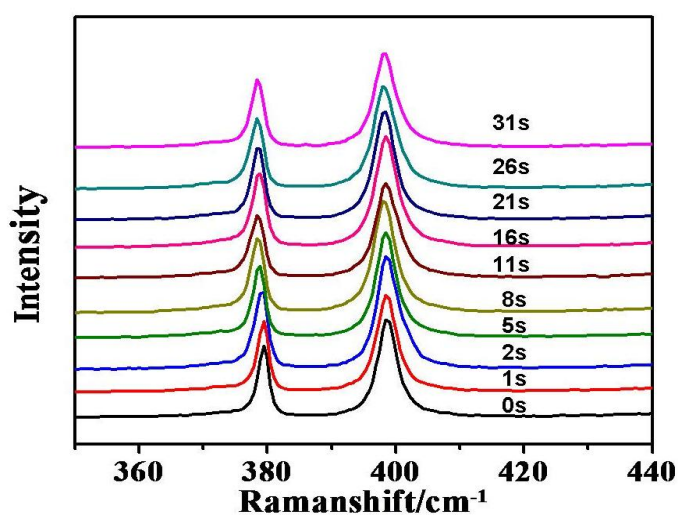

**Figure S3**. Raman spectra of monolayer MoS₂ after oxygen plasma irradiation with different time. No obvious changes have been observed, indicating the high quality of sample after plasma treatment.

**References**:
[1] Henkelman, G.; Uberuaga, B. P.; Jónsson, H. A Climbing Image Nudged Elastic Band Method for Finding Saddle Points and Minimum Energy Paths. *J. Chem. Phys.* **2000**, *113*, 9901
[2] Wang, J. L.; Ma, L.; Yuan, Q. H.; Zhu, L. Y.; Ding, F. Transition-Metal-Catalyzed Unzipping of Single-Walled Carbon Nanotubes into Narrow Graphene Nanoribbons at Low Temperature. Angew. Chem. Int. Ed. **2011**, *50*, 8041.